\documentclass[preprint,notitlepage,superscriptaddress,amsmath,amssymb,prl]{revtex4-2} 

\usepackage{braket}
\usepackage{miller}

\usepackage{graphicx}% 
\graphicspath{{Figures/}}
\usepackage[utf8]{inputenc}
\usepackage{multirow}
\usepackage{float}
\usepackage{bm}% bold math
\usepackage{verbatim}
\usepackage{xfrac}
\usepackage{array}
\usepackage{siunitx}
\usepackage{xspace}
\usepackage{comment}
\usepackage[textsize=scriptsize, bordercolor=black!50, backgroundcolor=black!10!white, linecolor=black!50]{todonotes}
\newcommand{\checkmeAl}[2]{{\color{teal}[Al]}{\color{red}#1}{ \color{blue}\ #2}}

\begin{document}
	\title{Probing Hyperbolic and Surface Phonon-Polaritons in 2D materials using Raman Spectroscopy}
 	\author{A. Bergeron}
	\affiliation{Génie physique, Polytechnique Montréal}
 	\author{C. Gradziel}
	\affiliation{Génie physique, Polytechnique Montréal}
	\author{R. Leonelli}
	\affiliation{Physics department, Université de Montréal}
	\author{S. Francoeur}
	\email{sebastien.francoeur@polymtl.ca}
	\affiliation{Génie physique, Polytechnique Montréal}
	
	\date{\today}
	%%%%%%%%%%%%%%%%%%%%%%%%%%%%%%%%%%%%%%%%%%%%%%%%%%%%%%%%%%%%

\begin{abstract}
The hyperbolic dispersion relation of phonon-polaritons (PhPol) provides high-momentum states, highly directional propagation, subdiffractional confinement, large optical density of states, and enhanced light-matter interactions. In this work, we use Raman spectroscopy in the convenient backscattering configuration to probe PhPol in GaSe, a 2D material presenting two hyperbolic regions separated by a \textit{double} reststrahlen band. By varying the incidence angle, dispersion relations are revealed. Raman spectra calculations confirm the observation of one surface and two extraordinary guided polaritons and matches the evolution of PhPol frequency as a function of confinement. Resonant excitation close to the excitonic state singularly exalts the scattering efficiency of PhPol. Raman spectroscopy of PhPol in non-centrosymmetry 2D materials does not require any wavevector matching strategies. Widely available, it may accelerate the development of MIR nanophotonic devices and applications.
\end{abstract}	

{
\let\clearpage\relax
\maketitle
}
%\maketitle
%\textit{Polaritons}
Strong coupling between an electromagnetic wave and a solid-state polar excitation yields hybrid quasiparticles known as polaritons \cite{Hopfield1958}. Owing to their mixed light-matter characteristics, polaritons express new properties and enable a unique level of control over light, matter, and their interactions. In this regards, 2D materials have been a prolific material platform for the development of polaritonic devices \cite{Basov2016}. 
%\textit{ph-Polariton}
Phonon-polaritons (PhPol) results from the coupling of polar transverse optical phonons with light \cite{Henry1965}. In strongly anisotropic nanostructures, they exhibit a hyperbolic dispersion relation from which high-momentum states become accessible, providing the means to achieve extreme energy confinement, large density of states enhancements, and sub-diffraction imaging in the mid-infrared region of the spectrum \cite{Dai2014,Caldwell2014} Along with the means to control their propagation \cite{Ma2018} and topological state \cite{Alu.Nature.2020}, their dispersion can also be dynamically shaped \cite{Taubner.NatureMaterials.2016}. These are a some of the few  functionalities that can be used to implement low-loss nanophotonic circuits in the infrared \cite{Kim.Jang.AdvOptMat.2020}. 

%\textit{s-SNOM}
Due to the large wavevector mismatch between free-space propagating photons and polaritons, special means such as prisms, gratings or nanostructures are required to couple PhPol to far field instrumentation such as light sources and detectors \cite{Luo2020}. In recent years, the exploration of PhPol has extensively relied on scattering-type scanning near-field optical microscopy (s-SNOM) \cite{Dai2014,Caldwell2014,Ma2018,Alu.Nature.2020,Taubner.NatureMaterials.2016}. This advanced near-field technique provides compelling images of propagating PhPol as a function of frequency, which reveal the $\omega-k$ dispersion relation. Similar information can be obtained with higher spatial but lower energy resolution using electron energy loss spectroscopy (EELS) \cite{Govyadinov.Hillenbrand.NatComm.2017}, a technique implemented within a scanning transmission electron microscope. 
%\textit{Materials}
 A large fraction of the recent literature on PhPol has focused on a few prototypical material systems, such as h-BN and $\alpha$-MoO$_3$.  If their low-loss, long-lived and tightly confined modes undoubtedly position them favorably, all 2D materials sustaining polar vibrational modes exhibit hyperbolic behavior that could further enable the development of infrared polaritonics. 
%\textit{Problematic}
The exploration of PhPol has so far relied on rather specialized and sophisticated instruments and has been limited to very few host materials with relatively high-frequency PhPol. 

%\textit{In this work...}
In this work, we demonstrate that Raman spectroscopy, just like s-SNOM and EELS, is an extremely powerful technique for studying PhPol. Owing to the relaxation of selection rules and the deep sub-wavelength confinement in thin samples, dispersion curves, 
confinement, and interactions with excitonic resonances can be studied in a backscattering configuration without the need from near-field or other wavevector matching strategies. %%%%%%%%%%%%%%%%%%%%%%%%%%%%%%%%%%%%%%%%%%%%%%%%%%%%%%%%%%%%%%%%%%%%%%%%%%%%%%%%%%%%%%%%%%%%%%%%%%%%%%%%

\begin{figure*}
    \centering
    \includegraphics[width=\textwidth]{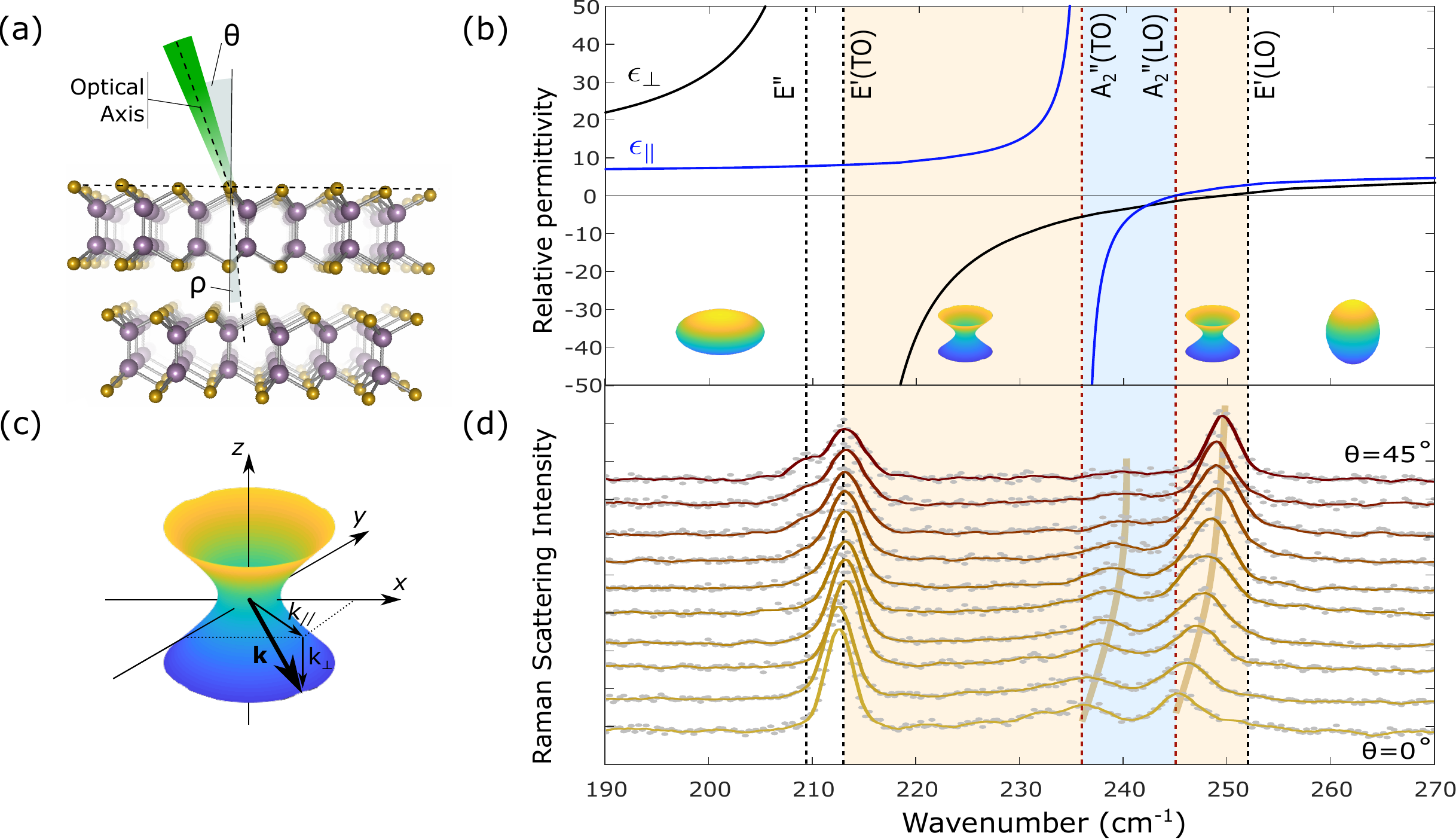}\
        \caption{(a) Crystal structure of $\epsilon$-GaSe and measurement configuration. $\theta$ controls the in-plane wavevector of the probed polaritons. (b) Permittivity for the two principal directions. The two type-II hyperbolic regions are shown in yellow. In between, there is a \textit{double} Reststrahlen (blue band). (c) Type-II hyperbolic dispersion of PhPol. (d) Raman backscattering from a \SI{650}{\nano\meter} thick sample as a function of $\theta$. Measured data are represented by the gray points, and solid lines are 5-point Savitzky-Golay filtered spectra. The \SI{532}{\nano\meter} excitation is $p$-polarized and $\theta$ is varied between 0 and \SI{45}{\degree} in \SI{5}{\degree} increments.}
    \label{F1}
\end{figure*}

For this demonstration $\epsilon$-GaSe is selected due to its strong polar resonances, nested Reststrahlen bands, and double type-II hyperbolic regions (Fig. \ref{F1}b). The $\epsilon$ polytype has an ABA stacking order, no inversion symmetry, and a D$^1_{3\rm{h}}$ space group \cite{Hoff1975,Hayek1973}. The crystal structure of $\epsilon$-GaSe is shown in Fig. \ref{F1}a) and a detailed description is provided in section S1 of Supplementary Information. This layered mono-chalcogenide has a gap at  $\SI{2.02}{\electronvolt}$ \cite{Meyer1973} is one of the best known nonlinear crystals for near- to far-infrared operation \cite{Chen2009}.  {As a result, both monolayer and bulk second-harmonic generation (SHG) have been shown to be more efficient than for other 2D materials \cite{Zhou2015a} with the exception of InSe, a closely related compound \cite{Hao2019}.}As discussed below, these nonlinear properties contribute to the Raman scattering efficiency from long-range polarization waves, making GaSe an excellent prototypical system for studying polaritons. 
%\textit{Raman Intro}

In the commonly used Raman backscattering configuration, conservation of momentum limits interactions with solid-state excitations with wavevector $k$ about twice that of the incident photon ($k\sim \SI{e5}{\per\centi\meter}$). As this is far above the light line $k\sim\omega/c$, polaritons lose their photonic character and become mostly mechanical. Hence, polariton scattering is typically observed, since their first discovery \cite{Henry1965}, in a near-forward scattering configuration for which typical wavevectors ($k\sim \SI{1e3}{\per\centi\meter}$) can be reached (see section S2 for more details). We demonstrate next that polaritons can also be most conveniently observed in backscattering configuration for thin ($d\ll \lambda_0 $) samples, and in some special conditions for thick ($d> \lambda_0$) samples, where $\lambda_0$ is the free-space polariton wavelength. 

%\textit{Angular dispersion of Raman modes}
    
Backscattering Raman measurements were performed on multiple thin exfoliated GaSe samples (see Methods) as a function of the sample tilt angle ($\theta$ in Fig. \ref{F1}a)). Strong angular dispersion of several Raman modes were observed in regions corresponding to the \textit{double} Reststrahlen bands ($\epsilon_\perp<0$, $\epsilon_\parallel <0$, blue band in Fig. \ref{F1}b)) and the upper type-II hyperbolic band ($\epsilon_\perp<0,\,\epsilon_\parallel>0$, yellow band). The data presented in Fig. \ref{F1}d) from a \SI{650}{\nano\meter} sample is used to identify these modes. At normal incidence ($\theta=0$\si{\degree}), the expected Raman-active zone-center modes A${_1^\prime}^1$ (non-polar, not shown), E$^\prime$(TO) (polar, shown) and A${_1^\prime}^4$ (non-polar, not shown) dominate the spectrum. By tilting the sample, excitations with non-zero wavevectors in the sample plane ($k_\parallel$) can be probed. Several important changes can be observed. First, the mode E$^{\prime\prime}$ appears right below E$^\prime$(TO). Forbidden at $\theta=0$, this non-polar and therefore purely mechanical mode becomes allowed and observable at incident angles larger than \SI{30}{\degree}. More interestingly, two relatively broad features can be readily identified, at all angles, in the region between 230 and \SI{255}{\per\centi\meter}. In contrast to all other modes, their energies do not match the tabulated frequencies of A$_2^{\prime\prime}$(TO), A$_2^{\prime\prime}$(LO) and E$^{\prime}$(LO). Furthermore, A$_2^{\prime\prime}$(TO) and A$_2^{\prime\prime}$(LO) are Raman-forbidden in all scattering configurations and E$^{\prime}$(LO) is, like E$^{\prime\prime}$, forbidden in a normal backscattering configuration and should only appear at large incident angle ($\theta>$ \SI{30}{\degree}). The angular dispersion of these two features is instrumental in further ruling out their non-polaritonic character.  As presented in section S3, the transverse extraordinary (Te) phonon connects E$^{\prime}$(TO) to A$_2^{\prime\prime}$(TO) and the longitudinal extraordinary (Le) phonon connects  A$_2^{\prime\prime}$(LO) to E$_2^{\prime}$(LO) \cite{Hoff1974}. Hence, the first dispersive branch located in the double reststrahlen (blue band) and evolving from \SI{236.4}{\per\centi\meter} ($\theta=\SI{0}{\degree}$) to \SI{241.6}{\per\centi\meter} ($\theta=\SI{45}{\degree}$) does not correspond to a phonon. The second branch, evolving from 247.8 to \SI{251.9}{\per\centi\meter}, is in a region where the Le phonon is expected, but the angular shift observed is almost 5 times more significant than that calculated from the lattice anisotropy. Indeed, the phonon dispersion implies a shift of at most \SI{0.9}{\per\centi\meter} \cite{Hoff1974,Hlinka2002}, as illustrated in Fig. SF2b). The observed shift of \SI{4.1}{\per\centi\meter} cannot be accounted for using mechanical phonons. 

%%%%%%%%%%%%%%%%%%%%%%%%%%%%%%%%%%%%%%%%%%%%%%%%%%%%%%%%%%%%%%%%%%%%%%%%%%%%%%%%%%%%%%%%%%%%%%%%%%%%%%%%

%\textit{Confined polariton branches}
    
We demonstrate that these dispersive branches originate from polaritons measured in a backscattering configuration by directly comparing the experimental spectra with the calculated Raman spectra. The model, fully detailed in Sections S3 to S9 of Supplementary Information, is applicable to any complex anisotropic multilayer structures. The Raman spectra is calculated from \cite{Mills1976,Sasaki1983} 

\begin{align}
    I_N(\omega)&= \frac{C(n_\omega+1)}{d} \cdot H^E_N(\omega,\theta) \delta{(\Delta\bm{q}_\parallel-k_\parallel)}\nonumber \\ &\left|(\hat{e}_i\cdot\widetilde{R}_N\cdot\hat{e}_s) \int_{-d/2}^{d/2}e^{i\Delta\bm{q}_\perp z}\langle E_N(z)\rangle {\rm d}z\right|^2,\label{E-Main}
\end{align}

\noindent where $N$ refers to the polariton normal coordinates: $To$ for transverse ordinary ( $To\perp(\hat{z},\bm{k})$), $Te$ for transverse extraordinary ($Te\perp(To,\bm{k})$), and $Le$ for longitudinal extraordinary ($Le\parallel\bm{k}$). $k$ is the polariton wavenumber, $C$ is an arbitrary constant, $n_\omega$ is the Bose-Einstein occupation factor, $d$ is the sample thickness, $\hat{e}_{i,s}$ is the polarization of the incident or scattered waves, and $\widetilde{R}_N$ is an effective scattering tensor taking into account the Raman tensor, the second order susceptibility tensor and the Fröhlich interaction tensor, defined in the polariton normal coordinates as described in Ref. \onlinecite{Irmer2013}, but generalized to uniaxial materials. $\langle E_N(z)\rangle$ is the time-averaged root mean square value of the polariton electric field at depth $z$. This field is calculated using a 4x4 transfer matrix formalism fully adapted to multilayered anisotropic media \cite{Passler2017d}. The Hopfield coefficient $H^E_N(\omega,\theta)$ represents the energy fraction of the polariton stored in the field. The polariton momentum along the surface $k_\parallel$ is controlled by the incident angle and the change in light momentum is given by $\Delta\bm{q}=\bm{q_i}-\bm{q_s}$.  The integral over the scattering depth ($z$) defines the directionality of the scattered light and will prove instrumental in understanding the origin of the polariton scattering in both thin and thick samples.  

\begin{figure*}
    \centering
    \includegraphics[width=\textwidth]{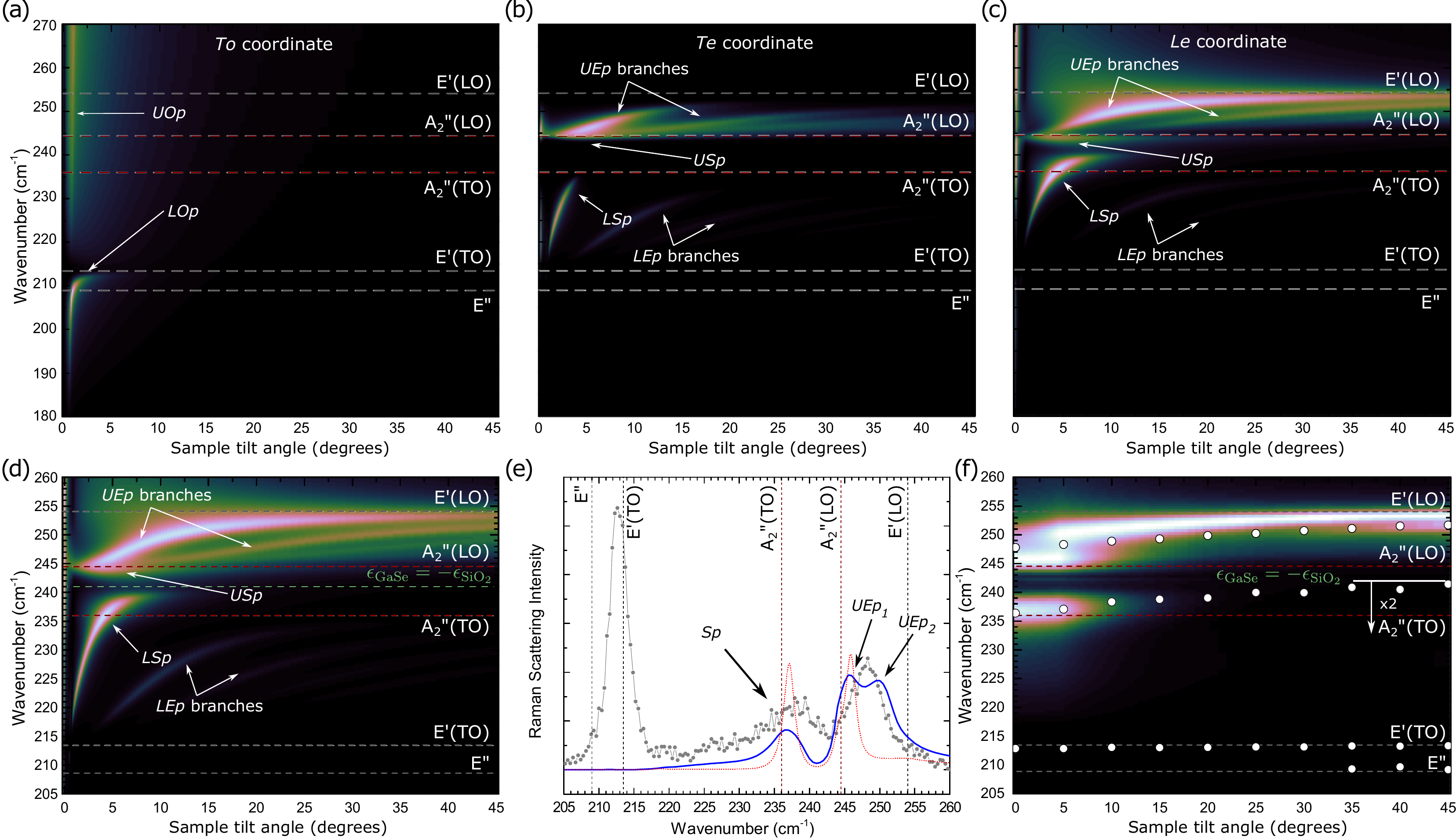}
    \caption{Calculated and measured Raman scattering intensity as a function of the sample tilt angle for a \SI{650}{\nano\meter}-GaSe sample on SiO$_2$/Si. (a-c) Calculated Raman spectra as a function of three normal coordinates ($To$, $Te$, $Le$). (d) Sum of the $Te$ and $Le$ coordinates. (e) Calculated (red), corrected for the acceptable angle (blue), and experimentally measured spectra for $\theta=\SI{5}{\degree}$ (gray). (f) Calculated (same as (d), but corrected for the finite acceptance angle) and measured polariton or phonon (E$^\prime(TO)$ and E$^{\prime\prime}$) dispersion. For all panels except (e), the intensity scale is logarithmic. The energies of the zone-center phonons are indicated by dashed lines. The green dashed line in (d) and (f) shows the lower surface polariton confinement criterion set by the permittivity of Si$O_2$.} 
    \label{F2}
    \end{figure*}

The calculated polariton scattering intensity is shown in \ref{F2}(a-c,d,e) as a function of sample tilt angle. The intensity resulting from each polariton normal coordinates is presented in Panels (a-c). Polaritons propagating at oblique angles can exhibit mixed character and branches are simply identified through their energy (lower or upper branch) and character (ordinary, extraordinary or surface). For the calculation of polariton scattering intensity and dispersion, it is found that GaSe polaritons are quite sensitive to their environment and it is particularly important to consider the whole Si/Si0$_2$/GaSe/air structure. As illustrated in Fig. SF3 of Supplementary Information for example, the polariton field distribution significantly extends into the silicon oxide and substrate, especially at lower incidence angles. 

Both the upper (UOp) and lower ordinary polariton (LOp) are highly dispersive and remain close to the light line with appreciable intensity only at angles below \SI{2.5}{\degree}. This can be explained by the fact that the ordinary polaritons are only weakly confined, as opposed to surface and hyperbolic polaritons. This high dispersion combined with the limited experimental angular selectivity (an important aspect that will be discussed further below) implies that ordinary branches may only appear as weak and very broad features (20-\SI{50}{\per\centi\meter}) that may not be resolved from the measurement noise. In this study, no ordinary polaritons have been identified and, since the $To$ coordinate does not mix with the other two, its contribution to Raman spectra will be ignored. 
    
Figure \ref{F2}(d) shows the Raman spectra calculated by summing the $Le$ and $Te$ coordinates. Three types of branches are identified: the lower and upper surface ($LSp$, $USp$) and ($LEp$, $UEp$) extraordinary polariton branches. The $LEp$ branch appears between E$^\prime$(TO) and A$_2^{\prime\prime}$(TO) at angles in the range between 4 and \SI{20}{\degree}. These branches have a relatively low scattering cross-section. Several $UEp$ branches are found between A$_2^{\prime\prime}$(LO) and E$^{\prime}$(LO). These polaritons contribute significantly to the Raman spectra at all angles. The two surface branches can be readily identified from the field distribution shown in Fig. SF3. The $USp$ is located at the GaSe/air interface. Its longitudinal coordinate dominates and its frequency is only slightly below that of A$_2^{\prime\prime}$(LO), but its Raman scattering cross-section is relatively low. In contrast, the lower surface branch is located at the GaSe/Si0$_2$ interface. It shows significant dispersion and scattering cross-section. It starts at the light line (not shown), crosses the energy of A$_2^{\prime\prime}$(TO) and extrapolates at high angles to the frequency determined by the surface mode confinement criterion ($\epsilon_\parallel(\omega_{Sp})\leq -\epsilon_{\rm{SiO}_2}(\omega_{Sp})$) indicated by the dashed green line at $\omega_{Sp}=\SI{241}{\per\centi\meter}$ in Panel (d) and (f). 

These calculated Raman spectra cannot be directly compared to the experimental data without first considering the angular selectivity of the Raman measurement system. The wide excitation and collection apertures typically used to increase spatial resolution limits the angular resolution required for resolving the dispersion relations. The relatively low numerical aperture of 0.19 of the system nonetheless corresponds to a collection angle of $\pm\SI{13}{\degree}$. Assuming a Gaussian beam profile, the acceptance angle $\Delta \theta$ at full width at half-maximum is \SI{15}{\degree}. This limited angular resolution can significantly affect the Raman spectra, especially at small angles where dispersion is important. This is illustrated in Fig. \ref{F2}e). At $\theta=\SI{5}{\degree}$, the features of the theoretical Raman spectra (red curve) do not satisfactorily match the experimental ones (gray points). However, taking into account the simultaneous contribution of multiple angles in a region of relatively high dispersion yields a calculated polariton Raman scattering spectra (blue curve) matching the experimental one. It is important to note that this calculated spectrum does not depend on any free adjustable parameter, as detailed in section S9, and can be used to identify the detailed origin of the Raman experimental features.

In Fig. \ref{F2}e), the unresolved experimental feature at \SI{247}{\per\centi\meter} results predominantly from the upper extraordinary branches ($UEP$) and, to a lesser extent, the upper surface polariton (USp). The two UEp both have significant $Te$ and $Le$ coordinates. The later makes an important contribution to the spectra despite the low angle $\theta=\SI{5}{\degree}$ due to the finite angular acceptance of the instrumentation. The experimental feature at \SI{238}{\per\centi\meter} corresponds to the lower surface polariton ($LSp$) located at the Si0$_2$ interface, which has a dominant $Le$ coordinate. Because of its significant dispersion over the angular acceptance at $\theta=\SI{5}{\degree}$, the upper surface polariton has an asymmetric lineshape with a pronounced low energy wing. At higher angles, this asymmetry is less pronounced due to the lower angular dispersion, as seen in \ref{F2}(d). Finally, on the low frequency side of the lower surface polariton, at about \SI{226}{\per\centi\meter}, is a small contribution from the lower extraordinary polariton ($LEp$) with a dominant $Te$ coordinate. This contribution may explain the Raman scattering intensity observed between 222 and 230, but this lower extraordinary branch has not been reliably resolved yet. 

Fig. \ref{F2}f) reports the calculated Raman spectra corrected for the finite angular acceptance of the system. The calculated upper extraordinary polaritons now appear as a broad features at low angles, evolving into a single narrower mode extrapolating slightly below E$^\prime$(LO) at high angles. The calculated lower surface polariton $LSp$ starts at 236 and slowly rises towards \SI{241}{\per\centi\meter}. The angular acceptance angle significantly affects the measured dispersion. Because of the intrinsically low intensity and relatively important LEp dispersion, they are barely seen even on this logarithmic scale. Panel (f) also reports the polariton modes experimentally measured. Here, data points represent the dominant frequencies identified using a Lorentz-lineshape analysis of the experimental spectra. Although such a simple lineshape is not expected, it nonetheless adequately identifies the main polariton modes and allows comparing calculated and experimental results over a broad angular range. We find that the experimental and calculated polariton modes are in excellent agreement, further confirming their identification as a guided and surface polaritons. As expected, the latter extrapolates, at high angles, exactly to the frequency determined by  surface confinement discussed above (green dashed line). Although the agreement between the calculated and experimental spectra is generally very good, the calculated intensity of the lower surface polariton drops off more rapidly than experimentally observed. This may be attributed to the high sensitivity of surface polaritons to the sample surface quality, roughness and contaminant, trapped charges on the SiO$_2$ surface, GaSe oxide, and the overall dielectric environment \cite{Prieur1975,Fali2019}.
%%%%%%%%%%%%%%%%%%%%%%%%%%%%%%%%%%%%%%%%%%%%%%%%%%%%%%%%%%%%%%%%%%%%%%%%%%%%%%%%%%%%%%%%%%%%%%%%%%%%%%%%

The integral in Eq. \ref{E-Main} reveals the configuration in which guided polaritons can be observed experimentally. In near-forward, the exponential term is slowly varying since $\Delta q_\perp = q_i \cos\theta_i - q_s \cos \theta_s\sim 0 $. Integrated over thick samples $d$, the near-forward selection rule $\Delta q_\perp=k_\perp$ is recovered. In backscattering, $\theta_s\sim \pi + \theta_i$ and $\Delta q_\perp\sim 2q_i $, resulting in rapid oscillations along $z$ and significantly reducing the scattering efficiency from  slowly-varying polariton fields. Hence, backscattering is precluded from bulk samples and only lattice phonon ($k\sim 2q_i$) can be measured. However, backscattering can be efficient if the polariton is strongly confined with a period that is comparable to that of the exponential term. For the specific case of GaSe, this is the case for samples thinner than \SI{1}{\micro\meter} (see S10). This further underscores the fact that the commonly understood association between polaritons and near-forward scattering geometries only applies to thick samples. For van der Waals materials where sub-micron confinement is straightforward, polaritons can be conveniently probed in a backscattering geometry. Despite this, Raman backscattering from weakly confined polaritons can, in some special conditions, be observed in samples as thick as \SI{70}{\micro\meter}. As demonstrated in S11, forward scattered Raman signals can be redirected towards the collection optics by a reflection at the back of the sample if the sample is transparent to the scattered light. Hence, care must be applied in the assignment of phonons in 2D materials, as weakly confined polaritons may  contribute to the backscattered Raman spectra for both thin and thick samples.

\begin{figure}[H]
    \centering
    \includegraphics[width=8.5cm]{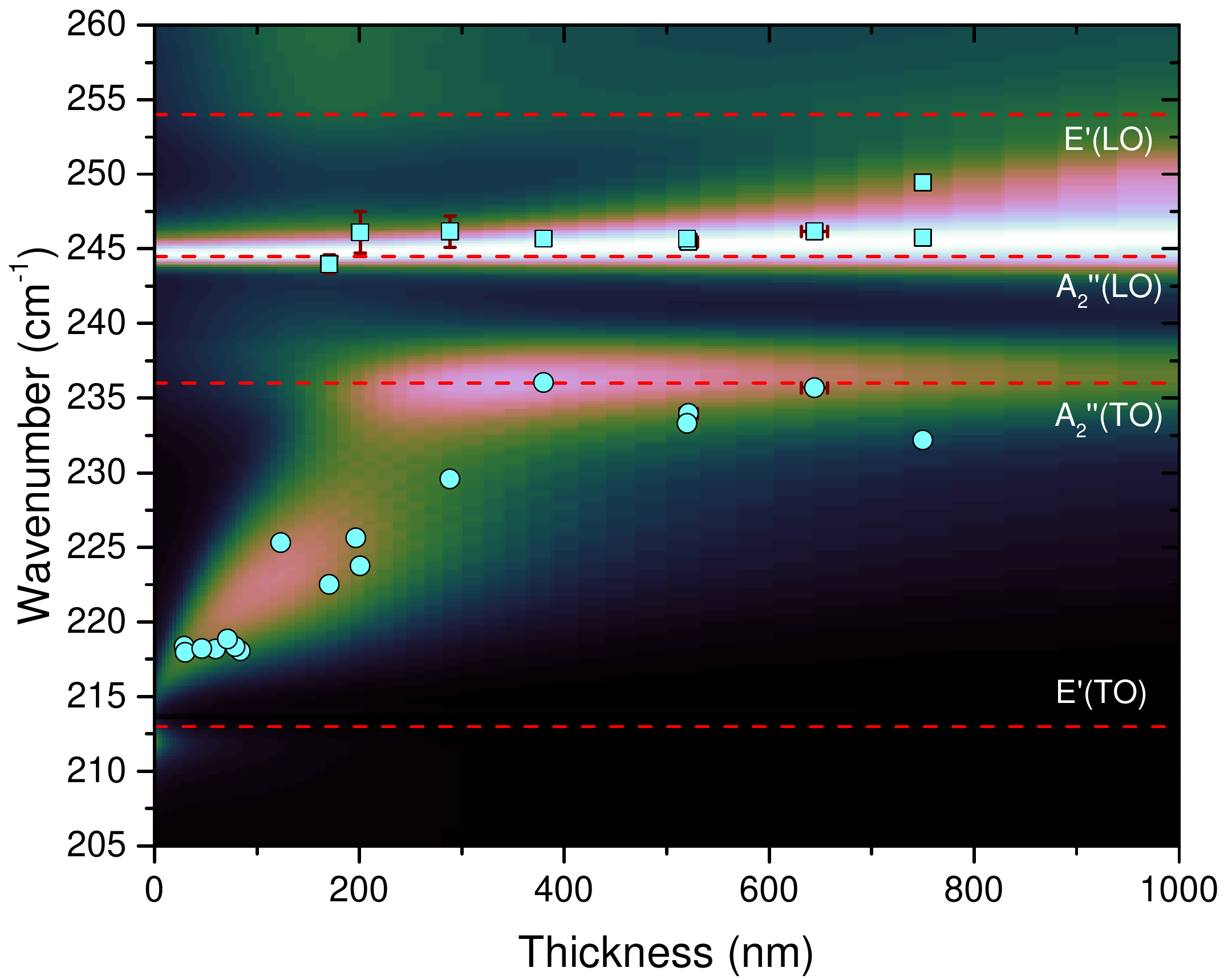}
    \caption{Raman spectra as a function of sample thickness for normal incidence and corrected for the finite acceptance angle. All spectra were individually normalized. As expected from the theory, the lower surface polariton ($LSp$, circles) and the upper extraordinary polaritons ($UEp$, white squares) dominate the experimental spectra. Around \SI{750}{\nano\meter}, the UEp are observed either as two resolved modes or a single wider one.}
    \label{F3}
\end{figure}

%\textit{Thickness dependency}
Engineering of the dispersion relations can be achieved through variation of the sample thickness and Raman spectroscopy can be used to investigate both guided and surface polariton frequency as a function of the sample or waveguide characteristics. Fig. \ref{F3} presents the calculated Raman spectra as a function of sample thickness. The spectra are corrected for the finite acceptance angle of the experimental measurement at $\theta=\SI{5}{degree}$. Each spectrum has been normalized according to its maximum value in order to illustrate the spectral weight rather than the thickness dependence of the intensity. As in Fig. \ref{F2}f), the calculated and experimental spectra are both dominated by the lower surface ($LSp$, circles) and upper extraordinary ($UEp$, squares) polaritons. Uncertainties in both frequency and thickness are shown, but for most data points uncertainties are smaller than the symbol size. From \SI{30}{\nano\meter} to \SI{750}{\nano\meter}, there is a strong agreement between calculated and experimental spectra. These results clearly demonstrate the relevance of Raman spectroscopy for the investigation of confined polaritons in 2D materials. For relatively thick samples, two guided UEp are sometimes observed, as expected from the calculation (see Fig. 2(e)). As demonstrated below, these two polaritons, identified as UEp1 and UEp2, are strongly exalted with excitation at the excitonic transition.

\begin{figure}[H]
\begin{center}
    \includegraphics[width=8.5cm]{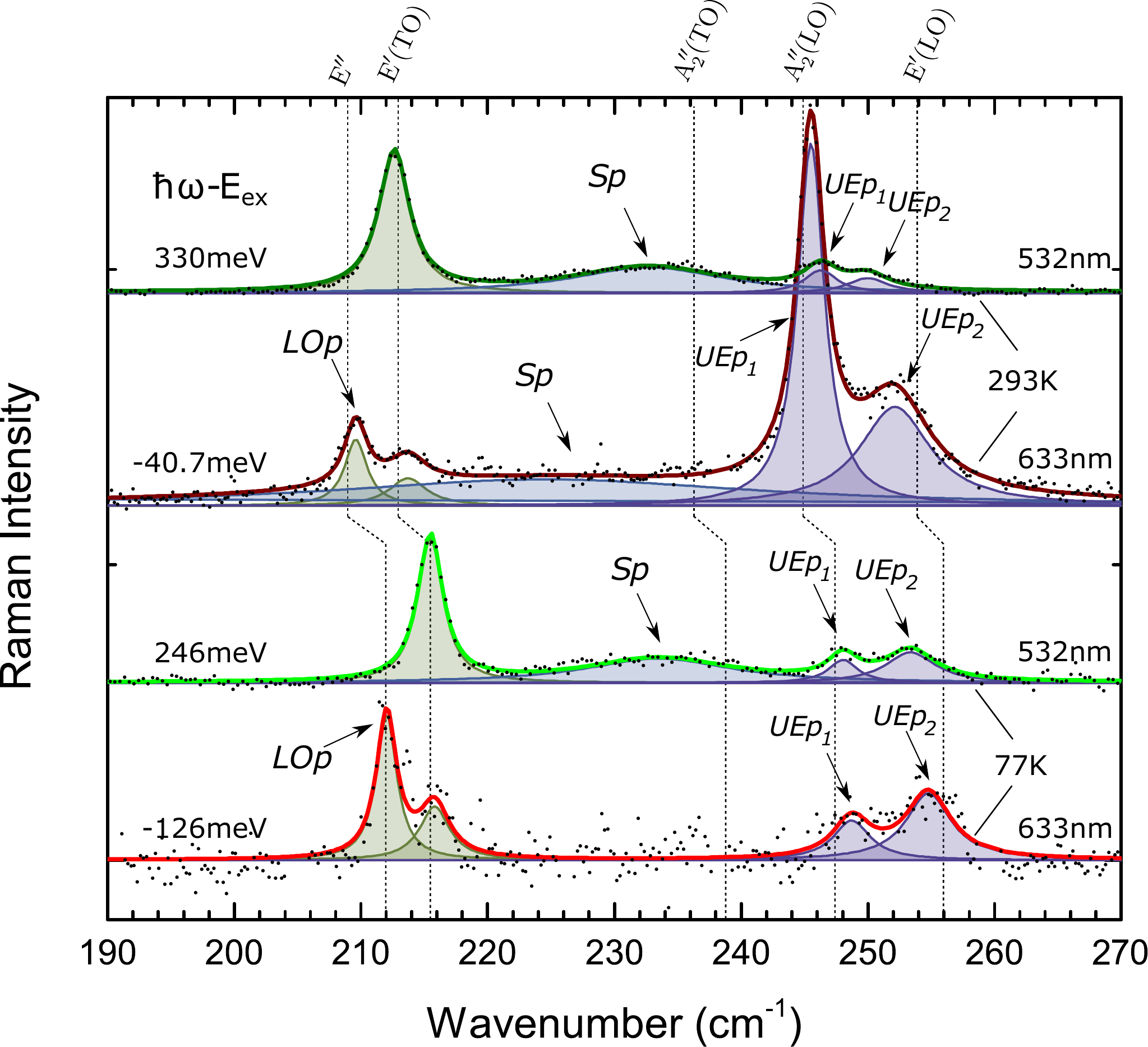}
    \caption{Raman spectra from 750-nm GaSe at 300 and \SI{77}{\kelvin}. Black symbols show the measured data, shaded regions illustrate the Lorentzian contribution from phonons (green) or polariton (purple), and the solid green and red lines represent the fitted spectra for \SI{532}{\nano\meter} and \SI{633}{\nano\meter} laser excitation, respectively. Lattice phonon energies are indicated by vertical dashed lines and the energy difference between the laser excitation and the 1$s$ exciton is shown on the left.}
    \label{F-Res}
\end{center}
\end{figure}

%\textit{Resonant Raman} 
Raman scattering efficiency is enhanced when either the incident ($\omega_i$) or scattered ($\omega_s$) photons are resonant with an electronic or excitonic transition ($\omega_X$), as the effective Raman tensor $\widetilde{R}$ presented in S6 become proportional $(\omega_{i,s}-\omega_X)^{-1}$. In addition, an additional $k$-dependent Fröhlich terms must be added and leads to a significant resonant enhancement of LO phonons for excitation energies near that of the $1s$ exciton \cite{Martin1971a, Martin1971b}. The relatively large exciton binding in bulk GaSe (19.2$\si{\milli\electronvolt}$) enables excitonic effects at 300 K \cite{Ferrer-Roca1999,Budweg2019}. Fig. \ref{F-Res} presents Raman spectra from a \SI{750}{\nano\meter}-GaSe sample measured as a function of temperature (300 and \SI{77}{\kelvin}) and laser excitation (532 and \SI{633}{\nano\meter}). To compare the effect of temperature and excitation wavelength, the spectra were normalized with respect to that of the totally symmetric and non-polar A$_1^{\prime 4}$ phonon. The 532 nm excitation exceeds the 1$s$-exciton by \SI{330}{\milli\electronvolt} (\SI{246}{\milli\electronvolt} meV) at \SI{300}{\kelvin} (\SI{77}{\kelvin}). Far away from resonance at both temperatures, the two spectra are almost identical besides the temperature-induced frequency shifts: they both show two first and second upper extraordinary polaritons ($UEp$) and the surface polariton ($Sp$). Temperature alone does not qualitatively affect the polariton Raman spectra. In contrast, the \SI{633}{\nano\meter}-excitation is only \SI{-40.7}{\milli\electronvolt} away from the  1$s$-exciton at \SI{300}{\kelvin}. This near-resonance induces several striking changes. First, both $UEp$ are significantly exalted, since the Frölich term is particularly important for polaritons with a strong longitudinal coordinate. At frequencies where the surface polariton is expected, only a very broad continuum is found. Although the lack of a well-defined lineshape prevents a clear identification, the highly dispersive lower extraordinary polaritons ($LEp$) are expected in this spectral region and may contribute to the observed Raman spectra. Finally, a new polariton appears slightly above the purely mechanical E$^{\prime\prime}$ phonon, which is forbidden in this normal backscattering geometry ($\theta=0$) and should only be observed at very high angles ($\theta\geq\SI{35}{\degree}$). As shown in Fig. \ref{F2}(a) for a slightly thinner sample, the lower ordinary polariton ($LOp$) is expected in this energy range and has already been identified as such in Ref. \cite{Bianchini2012} through forward scattering signals from bulk GaSe.  The concomitant effects of low dispersion and resonant enhancements result in a well-defined Raman mode. 

Cooling the sample to \SI{77}{\kelvin} moves the $1s$-exciton further from the laser excitation. As expected, the Fröhlich contribution of longitudinal modes is quenched and the intensity from both $UEp$ modes is back to non-resonant levels. In contrast, $LOp$ remains exalted even \SI{-126}{\milli\electronvolt} below the $1s$-exciton. In fact, this mode has been observed with excitation as far as \SI{410}{\milli\electronvolt} below the exciton \cite{Bianchini2012}, indicating that the resonance process involves a much slower decay than that expected from a pure Fröhlich interaction. The resonant Raman results offer an interesting insight into the coupling between the excitonic and polaritonic states, and provide an interesting window into the rich electron band structure and exciton dynamics of the host material. Furthermore, the extreme enhancement of the Raman cross-section near resonance enables a closer study of polaritonic effects which could not be observed away from resonance due to weaker signals.

%\textit{Conclusion}
We have demonstrated that Raman spectroscopy is a powerful tool for the study of PhPol and, due to the lack of wavevector matching requirements, conveniently implemented as well. 
It allows mapping the low energy of phPol (5-\SI{150}{\milli\electronvolt}) into the visible where single-photon sensitivity is readily available. Albeit Raman spectroscopy is generally associated with a weak cross-section, resonant excitation significantly enhances the scattering efficiency and allows probing excitonic states. Naturally, Raman spectroscopy of polar excitations is restricted to non-centrosymmetric crystals. This includes 3R- and Td-TDMC for examples as well as odd number of layers h-BN and 2H-TDMC. It also includes crystals for which inversion symmetry is broken by a discontinuity or a perturbation such as an external field. Almost universally available in 2D material research labs, backscattering Raman spectroscopy may prove to be a strategic technique for the study of phPol and may accelerate the development of a wider variety of polaritonic materials and devices.

\section*{Methods}
\subsection*{Sample preparation and protection} 
Thin samples were first cleaved from Bridgman grown GaSe crystals to expose pristine, optically clear and oxidation-free samples and then transferred onto a PDMS pad. Samples were mechanically exfoliated \cite{Castellanos-Gomez2014} on a \SI{300}{\nano\meter} thermally grown silicon oxide. Thickness of thin and thick samples was determined from tapping-mode atomic force microscope or from broadband transmission interferometry using two incident angles. To avoid sample degradation and photo-oxidation \cite{Bergeron2017}, exfoliation and AFM measurements were done in a dry-nitrogen filled glovebox and angle-resolved Raman measurements were performed using a vacuum optical cell (\SI{1e-5}{\milli\bar}) with low-birefringence windows.

\subsection*{Backscattering Raman experiments}
A typical modular Raman bench was used. Laser excitation was provided by a HeNe (\SI{633}{\nano\meter},
$\leq$\SI{1.5}{\giga\hertz} linewidth) or a diode-pumped Nd-YAG (\SI{532}{\nano\meter}, \SI{1}{\mega\hertz} linewidth) laser. All spectra were measured using a 532-nm laser, except for those indicated in Fig. 4. An underfilled objective with a working distance of \SI{9.1}{\milli\meter} provided an effective numerical aperture of 0.19. For polarization control, both the excitation and scattered beam passes through a zero-order $\lambda/2$ waveplate. Parallel or orthogonal detection configuration is selected by adding a polarizer on the path of the scattered beam, followed by a Raman filter and a detection system (spectrometer and CCD camera) providing a spectral resolution better than. Energy calibration was based on the omnipresent non-polar modes A$_1^1$ and A$_1^4$. These do not exhibit directional dispersion and are reliable energy markers. Typically, Raman spectra were acquired using a total exposure time of \SI{15}{\minute} using a fluence of the order of \SI{200}{\micro\watt\per\micro\meter\squared}. All measurements are free from thermal effects. 

%\clearpage
%\bibliography{library}
\bibliography{Biblio1,Biblio1a,Biblio2,Biblio2a,Biblio3}

\section*{Acknowledgments}
This work was supported in part by the Natural Sciences and Engineering Research Council of Canada (NSERC) and the Canada Foundation for Innovation. A.B. received a FRQNT scholarship. 

\section*{Author contributions}
A.B. and C.G. have prepared samples, performed experimental measurements (AFM, optical microscopy, Raman). A.B. has performed all calculations. R.L. and S.F have supervised the project. S.F. and A.B. wrote the manuscript with the help of C.G. and R.L.

\section*{Competing financial interests}
The authors have no competing financial interests. 

\begin{comment}
%\section*{To DO}
\begin{enumerate}
    \item \checkmeAl{What is the enhancement of UEp1 and UEp2 under quasi-resonant excitation}{?}
%    \item Calculate the group velocity
   
  %  \item Its  second-order nonlinear coefficient ($d_{22}$) is more than one order of magnitude above those of common nonlinear crystals \cite{Nikogosyan2005}.
%    \item Discuss the efficiency of Raman with respect to the non-linear coefficients. 
 %   \item Extra text- The study of phonon-polaritons requires overcoming the momentum mismatch between free-space propagating fields and guided phonon-polaritons.
    \item Concernant la nomenclaure pour UOp et LOp dans la figure 2. Message d'Alaric: Pour le 4, je sais que je nommais ces branches $To$ dans ma thèse, mais je dois me familiariser avec la nomenclature de l'article et me replonger dans la soupe pour me souvenir de l'origine de ces deux branches. C'est possible que ce soient des branches extraordinaires (E).
\end{enumerate}
\end{comment}

\end{document}